# Non-volatile switching in graphene field effect devices

T.J. Echtermeyer[a][1], M.C. Lemme[a], M. Baus[a], B.N. Szafranek[a], A.K. Geim[b], H. Kurz[a]

[a]Advanced Microelectronic Center Aachen (AMICA), AMO GmbH, 52074 Aachen, Germany

[b]Manchester Centre for Mesoscience and Nanotechnology, University of Manchester, UK

## Abstract

The absence of a band gap in graphene restricts its straight forward application as a channel material in field effect transistors. In this letter, we report on a new approach to engineer a band gap in graphene field effect devices (FED) by controlled structural modification of the graphene channel itself. The conductance in the FEDs is switched between a conductive "on-state" to an insulating "off-state" with more than six orders of magnitude difference in conductance. Above a critical value of an electric field applied to the FED gate under certain environmental conditions, a chemical modification takes place to form insulating graphene derivatives. The effect can be reversed by electrical fields of opposite polarity or short current pulses to recover the initial state. These reversible switches could potentially be applied to non-volatile memories and novel neuromorphic processing concepts.

**Index Terms**

*graphene, field effect transistor, memory, MOSFET, non-volatile, switch*

**Not**e

---

[1] phone: +49 241 8867 229, email: echtermeyer@amo.de

# I. INTRODUCTION

Graphene has been demonstrated to posses remarkable intrinsic electronic properties that include carrier mobilities exceeding 200,000 cm2/Vs and a micrometer-scale mean free path at room temperature [1][2][3]. As a consequence, one of the most interesting applications in nanoelectronics is based on the use of graphene as channel material for field effect transistors (FET) [4]-[10]. However, the minimum conductance of macroscopic graphene even at the neutrality (Dirac) point, where no carriers are nominally present, results in low $I_{on}/I_{off}$ ratios by far insufficient for CMOS-type applications. Indeed, the best ratio achieved to date with a top-gated graphene FET at room temperature is only six [9]. A possible escape route is the use of graphene nanoribbons (GNRs) narrower than 5 nm, which in theory will have a band gap larger than 500 meV [11][12]. Very recently, these predictions have been verified experimentally using either random methods similar to typical carbon nanotube processes [13], or by deliberately overetching pre-defined structures [14]. Nonetheless, reliable top-down fabrication of such structures is not possible even with state-of-the-art nanolithography tools. The smallest GNR-FETs fabricated with controlled top-down lithography demonstrated so far have shown widths of ~10 nm [1][7][15], with electrical data available for ~30nm GNRs.

In this letter, we report on graphene-based switches in a field effect transistor configuration that rely on field-induced chemical modification of graphene's crystalline structure. $I_{on}/I_{off}$ ratios of over $10^6$ are achieved in these field effect devices (FEDs, we use this notion to indicate a different operational mechanism with respect to the conventional FETs) at room temperature.

## II. Experiment

The graphene FEDs have been fabricated by exfoliation of graphite on top of a silicon substrate with 300 nm of silicon dioxide ($SiO_2$) [16]. A 20 nm layer of silicon oxide ($SiO_x$) has been evaporated on top of graphene as a top-gate dielectric and a 40 nm tungsten film has been sputtered as source, drain and top-gate electrodes (for details see [4][8]). In a separate experiment, a 20 nm SiO$_2$ dielectric has been deposited by chemical vapor deposition (CVD) at 425°C. A schematic and an optical micrograph of the devices are shown in Figure 1a and 1b. An HP 4156 semiconductor parameter analyzer has been used for electrical measurements. More than ten samples have been fabricated, all of which exhibit the switching effect described below.

## III. Results and Discussion

Fig. 2a shows the top-gate transfer characteristic of a graphene FED with 2 µm channel width and 4 µm gate length measured in ambient conditions, with its Dirac point at $V_g = 0$ V and, as expected, with a limited $I_{on}/I_{off}$ ratio of ~1.5. This is a typical $I_{ds}/V_{tg}$ characteristic of a graphene field effect transistor (compare e.g. [4]-[8], [17]) and translates to a channel resistivity of 5 kΩ (labeled "on-state" in Fig. 2b) In this measurement, the top-gate voltage has been swept back and forth between $V_{tg} = -4$ V and $V_{tg} = 4$ V and the back-gate voltage has been kept constant at $V_{bg} = -40$ V.

In a consecutive measurement, the range of the top-gate voltage sweep has been extended above a certain critical value. When starting the sweep at $V_{tg} = -5$ V, the drain current drops by over seven orders of magnitude into an insulating state of the proposed graphene FED ("off-state"). However, this effect is reversible: as the voltage is swept starting from $V_{tg} = -5$ V towards $V_{tg} = 5$ V (Fig. 2b, filled circles), the device remains insulating for negative voltages but recovers almost to its initial on-state current for $V_{tg} > 0$ V. A low gate leakage current excludes breakdown of the gate oxide as the responsible mechanism (Fig. 2b, hollow circles). The inset in Fig. 2b shows the transfer characteristics of a device with a 20nm CVD SiO$_2$, which exhibits the same qualitative switching behavior despite a different dielectric is used.

The observed switching of the channel resistivity is attributed to a chemical modification of the graphene, induced by the electrostatic gate-field. Two examples for such modified graphene are graphane and graphene-oxide. Graphane is a derivative with hydrogen (H+) atoms attached in sp$^3$ configuration and a band gap of 3.5 eV [18]. Graphene-oxide is a graphene sheet with a large amount of hydroxyl (OH-) or similar groups attached to its surface and it is insulating at room temperature [19][20]. We suggest that in our experiment, water is split into H+ and OH-, which then attach to the graphene surface and open a band gap. However, further experiments in this letter are aimed at the application of the effect in a solid state device, while a detailed discussion of the underlying mechanisms and specific experiments including electrochemical gating are reported elsewhere [21].

Utilizing the observed switching effect in future electronic applications obviously requires reliable ways of controlling the switching and cycling it numerous times. Our approach includes the field induced switch-off as shown in Fig 2b, but replaces the field induced switch-on with short current pulses of 80 µs in length (limited by the experimental setup). Fig. 3 shows a time dependent measurement of the channel resistivity of a graphene FED. The top and back-gate voltages have been kept constant at $V_{tg}$ = -5.5 V and $V_{bg}$ = 0 V. Resistivity increases by over 6 orders of magnitude to ~$5 \times 10^{11}$ Ω and remains at this value. After reaching the off-state, a top-gate voltage of $V_{tg}$ = 0 V has been applied and the resistivity remains in the off-state which shows the device's non-volatile characteristic. Then a current pulse of 50µA has been applied for 80 µs, after which the channel resistivity recovers fully. Applying $V_{tg}$ = -5.5 V has turned the device off again. While the DC voltage compliance has been set to 8 V for the current pulse, the actual AC voltage can be much higher, but could not be measured with the DC unit.

Fig. 4 shows the range of the resistivity change between the on-state and the off-state (log ($I_{on}/I_{off}$)) for two devices which were both cycled eight times. Here, a top-gate voltage of $V_{tg}$ = -4.5 V has been applied to turn off the devices and single 50 µA / 80 µs current pulses have been used to restore the on-state. Device A displays a spread in the modification of the channel resistivity between three and eight orders of magnitude, compared to five to almost seven orders of magnitude in device B. The observed spread can be interpreted as intermediate states of resistivity caused by different levels of chemical modification of the graphene. In an additional experiment, a device has been switched to the off-state and has then been left in this condition in the clean room for two days, before returning it to the on-state with a standard current pulse. This hints at reasonably large retention times of the effect. Even though substantial further investigations are required to improve the switching reliability and -times and to identify potential storing times and cyclability, our experiments clearly demonstrate the potential of graphene field effect devices as non-volatile switches. In fact, the observed resistivity changes of more than six orders of magnitude are much larger than in "resistive" switches reported recently [22]. In addition, multistage logic seems possible, if resistance values in graphene FEDs as shown in Fig. 4 can be controlled reliably.

## IV. Conclusion

In this letter, we have reported on the modification of channel resistivity in a graphene field effect device by over six orders of magnitude. The general device structure is identical to conventional silicon-on-insulator and graphene MOSFETs and the devices may potentially be seen as candidates for future non-volatile memory applications. Our experiments show good cyclability and reset times of 80 µs. Better understanding of the involved processes should lead to improvements in switching times and reliability in the future, including finding a reliable source for the species involved in the switching mechanism. In the long run, the observed intermediate states with ample margins may be applicable to neuromorphic processors and networks.

**Acknowledgment**

The authors would like to thank J. Bolten and T. Wahlbrink for their e-beam lithography support. Financial support by the German Federal Ministry of Education and Research (BMBF) under contract number NKNF 03X5508 ("Alegra") is gratefully acknowledged.

**FIGURE CAPTIONS**

Fig. 1: a) Schematic of a double-gated graphene field effect device (FED) used in the experiments. b) Optical micrograph of several FEDs fabricated from one graphene flake.

Fig. 2: a) Typical top-gate transfer characteristics of a graphene FED with $SiO_x$ dielectric. b) Top-gate transfer characteristics including chemical switching of the graphene FED with $SiO_x$ dielectric and gate leakage current. Inset: Transfer characteristics of a device with CVD $SiO_2$ gate dielectric.

Fig. 3: Resistivity of a graphene FED over time, including three switching events (on -> off -> on -> off).

Fig. 4: Resitivity change in orders of magnitude for repeated switching from the on- to the off-states of two graphene FEDs.

**Fig. 1**

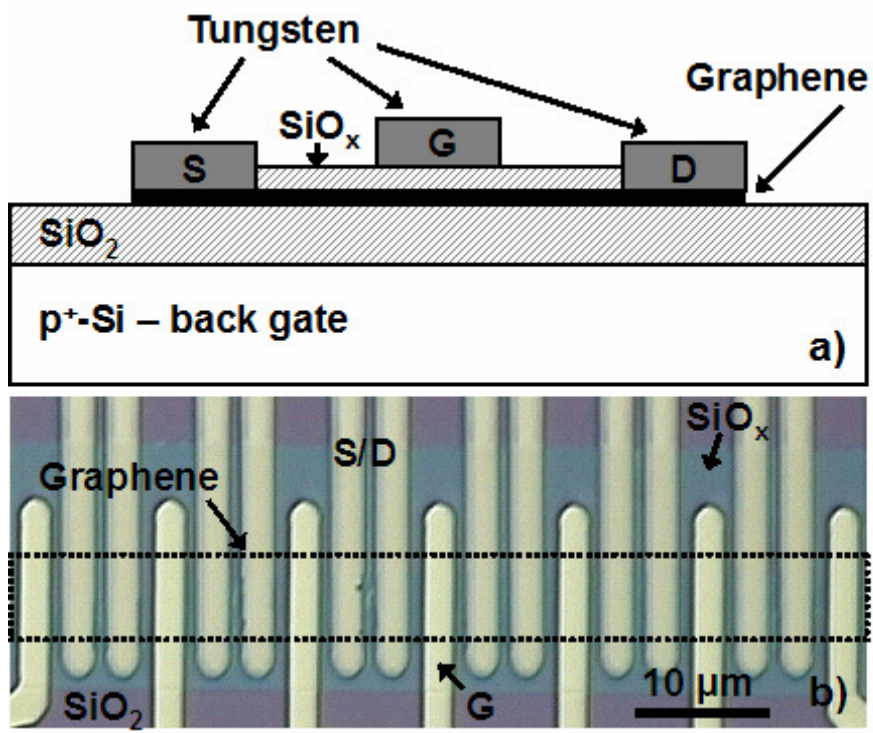

**Fig. 2**

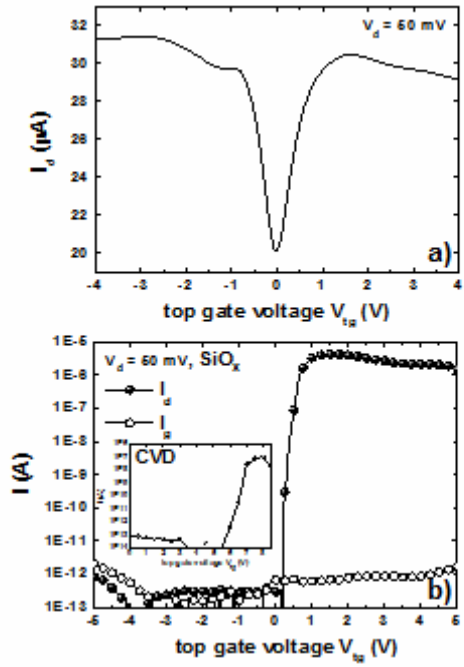

**Fig. 3**

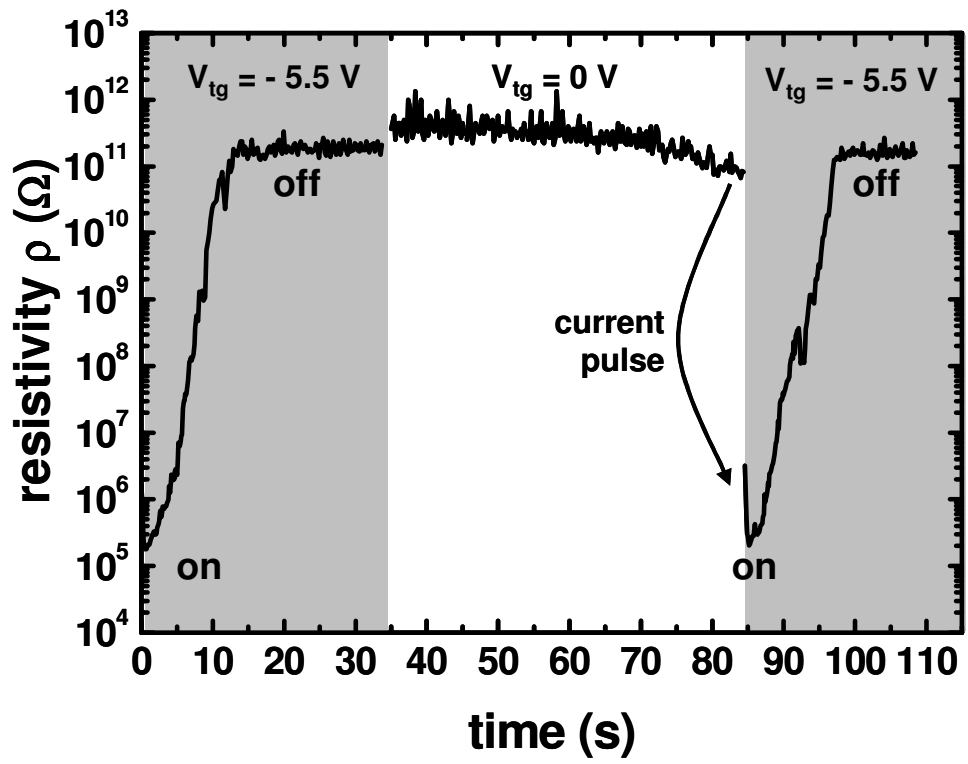

**Fig. 4**

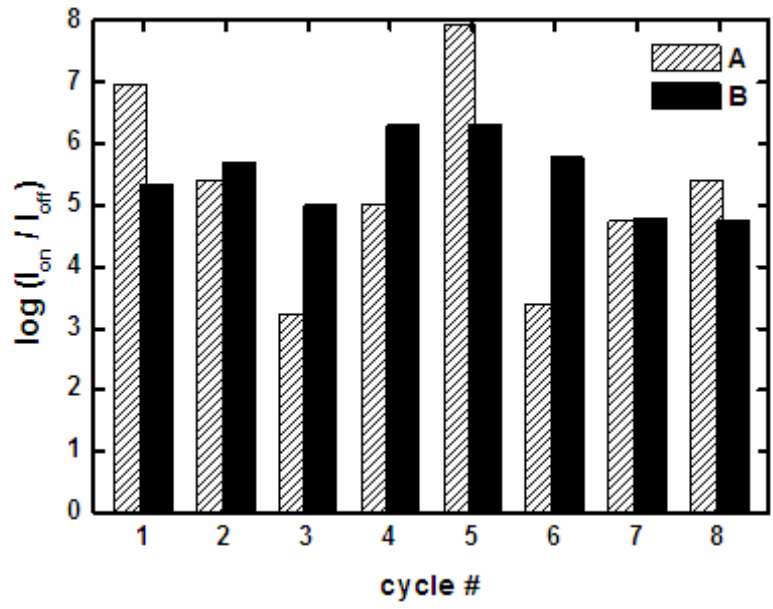